\documentstyle[12pt]{article}
\begin{document}
\newcommand{\be}{ \begin{equation} }
\newcommand{\ee}{ \end{equation} }
\newcommand{\bi}{ \bibitem }
\newcommand{\muh}{ \hat{\mu} }
\newcommand{\rd}{ \mbox{\rm d} }
\newcommand{\re}{ \mbox{\rm e} }
\renewcommand{\floatpagefraction}{0.8}
\begin{titlepage}
\begin{flushright} CERN-TH 6407/92 \end{flushright}
\vspace{1.5cm}
\begin{center}
{ \LARGE Scaling topological charge \\ in the $CP^3$ spin model}
\vspace{1.5cm}

{\large Ulli Wolff \\
CERN, Theory Division,\\
CH-1211 Gen\`eve 23, Switzerland}
\date{}
\end{center}
\vspace{1cm}
\thispagestyle{empty}
\begin{abstract}\normalsize
The $CP^3$ spin model is simulated at large correlation lengths in
two dimensions. An overrelaxation algorithm is employed which yields
reduced critical slowing down with dynamical exponents $z\approx 1$.
We compare our results with recent multigrid data
 on the massgap $m$ and the spin susceptibility
and confirm  the absence of asymptotic scaling.
As a new result we find scaling for
the universal topological susceptibility
with values extrapolating  to $\chi_t / m^2  = 0.156(2)$
 in the continuum limit.
\end{abstract}
\vspace{2cm}
\begin{flushleft} CERN-TH 6407/92 \\ March 1992 \end{flushleft}
\end{titlepage}
\section{Introduction}
With the discovery of Monte Carlo simulation algorithms with reduced
critical slowing down it has become possible to nonperturbatively study
a number of
two dimensional lattice field theories close to the continuum limit.
As the  ratio between physical length scales and the cutoff length reaches
values of O(100), these high precision simulations obtain a completely
new quality
compared with what could be done before.
Cluster algorithms \cite{cluster}
are particularly efficient, but so far their success
has remained restricted to $O(n)$ invariant
vector models as far as continuous field manifolds are concerned.
Recently multigrid techniques \cite{MGO3} succeeded in extending the class of
models accessible to such
high precision study to include the $CP^3$ and $SU(3) \times SU(3)$
systems with standard action \cite{HM1}.
In parallel with these nonlocal methods overrelaxation (OR) techniques have
been
developed for practical applications to spin systems \cite{ORO2,ORO3}.
Although OR is expected to only reduce critical slowing down to dynamical
exponents $z \approx 1$, thus producing a gain in efficiency proportional to
the
correlation length, it already makes a large difference in practice.
Being a local algorithm it can be implemented very efficiently,
with no overhead, and it is actually superior to multigrid for many simulations
at intermediate correlation lengths of O(10). In this letter we present OR
results on the $CP^3$ model. The purpose is on the one hand to offer an
independent check on results obtained with the new multigrid technique
in \cite{HM1} and on the other hand to supplement these data by values
of the topological susceptibility, a particularly interesting observable
in $CP^n$ models\footnote{ During our numerical work we received a copy
of \cite{HM2} where \cite{HM1} is refined by higher statistics,
and the topological charge is studied, too. The results
are now in slight conflict with ours, as we discuss in sect.3
}. In sect.2 we define the model and
present the OR algorithm used.
The physical results extracted for the $CP^3$ model
are discussed in sect.3.

\section{$CP^n$ spin model and overrelaxation}

The spins of $CP^{n-1}$ models live on the sites $x$ of a hypercubic
$L\times L$ lattice
and consist of rays in complex $n$ dimensional space. They can be labelled
by unit vectors $z(x)$ entering into the standard action only through
1-dimensional projectors
\be
P(x) = z(x) z^{\dagger}(x).
\ee
The partition function then reads
\be
Z=\int \prod_x \rd^n z(x) \, \rd^n z^{\dagger}(x)\,
\delta(|z(x)|^2-1) \, \exp\left\{2 \beta \sum_{x\mu}
\mbox{tr} [ P(x) P(x+\muh) ]\right\}\, ,
\ee
where $\mu =0,1$ and $\muh$ are unit vectors in the
corresponding lattice directions.
The fundamental correlation function is given by
\be \label{C=}
C(x) = \left\langle \mbox{tr} [P(x) P(0)] \right\rangle -1/n\, ,
\ee
and the magnetic susceptibility $\chi$ is constructed
by summing the argument of $C$ over all sites.
We employ periodic boundary conditions, and the mass
$m$ or correlation length $\xi=m^{-1}$ are defined through the
temporal decay of $C$ at zero spatial momentum
[we use lattice units putting the lattice spacing $a=1$].

Topological quantities in the $CP^{n-1}$ model are associated with
the ``composite'' $U(1)$ gauge field
\be
U_{\mu}(x)= \frac{z^{\dagger}(x) \cdot z(x+\muh)}
{|z^{\dagger}(x) \cdot z(x+\muh)|}\, .
\ee
As one forms plaquette fields $U_p$ from it and $F_p$ with
\be
U_p = \re^{i F_p} \, , \, -\pi < F_p \le \pi \, ,
\ee
the topological charge
\be
Q=\frac{1}{2\pi} \sum_{p} F_p
\ee
is integer, since the product over all $U_p$ is unity on the torus.
While the average of $Q$ vanishes for symmetry reasons (which we monitored),
the topological susceptibility
\be
\chi_t = \langle Q^2 \rangle / L^2
\ee
is expected to be a nontrivial physical quantity of length dimension $-2$.

Local updates of a $CP^{n-1}$ spin $z$ at some site $x$ have to sample the
local Boltzmann factor
\be \label{lB}
p(z) \propto \exp( z^{\dagger} M(x) z),
\ee
where the hermitean $n\times n$ matrix $M$ is given in terms of the
nearest neighbors,
\be
M(x)= 2 \beta \sum_{|y-x|=1} P(y).
\ee
The decisive component in our realization of OR are  microcanonical moves
of individual spins which leave (\ref{lB}) unchanged. As has already been
observed in \cite{JW}, this can be achieved, if a normalized
 eigenvector $\psi$ of
$M(x)$ is known, by reflecting
\be \label{newz}
z \rightarrow -z + 2 \psi \psi^{\dagger} \cdot z \,.
\ee
In the $CP^1$ model, which coincides with the $O(3)$ model, the standard
microcanonical reflections \cite{ORO3} are reproduced for either choice
of the two possible eigenvectors of $M$. For the case $n=4$, which we study
here, we use the eigenvector belonging to the largest eigenvalue, as that one
is easy to obtain.

In the practical realization of hybrid OR we perform
$N$ microcanonical sweeps with checkerboard ordering followed
by one standard ergodic Metropolis sweep. In the latter, we move the $z$
spins with $U(2)$ rotations of all possible pairs of components
of $z$. The eigenvectors of the $M(x)$ are simply constructed approximately
by multiplying $M$ a few times on a start vector preconditioned to the
sum of the nearest neighbor $z(y)$ spins.
Then the reflected spin (\ref{newz}) is first considered as an update
proposal, which is followed by an accept/reject step taking into account
small changes in energy due to the approximation.
With four multiplications we
 find for all our simulated $\beta$ values
 between 96\% and 99.7\% acceptance (growing with $\beta$),
which makes this step practically microcanonical.
Note that in this way the remaining small error in the eigenvectors does
not lead to any systematic bias in the simulation.

The resulting integrated autocorrelation times
as functions of the correlation length $\xi$ are summarized in Fig.1.
Errors are displayed but mostly fall within the symbols.
\begin{figure}
\vspace{11cm}
\caption{Integrated autocorrelation times
in single sweeps for energy $E$,
spin susceptibility $\chi$, and topological susceptibility $\chi_t$.}
\end{figure}
%
We employed the tuning strategy of taking the number of microcanonical
sweeps $N$ roughly proportional to $\xi$. This was found to be optimal
 both in an extensive numerical study of the $O(3)$ model
\cite{ORO3} and in a rigorous
analysis of the Gaussian free field model \cite{UWOR}.
More precisely, in the order of growing $\xi$, we took
$N=1,3,6,9,12,25$.
All $\tau_{\mbox{\scriptsize int}}$ refer to single sweeps, and we did
not distinguish OR and Metropolis here. In fact, close to the continuum limit,
only the microcanonical OR steps, which are faster, are important.
The dynamical exponents $z$ shown in Fig.1
result from fits to the four largest $\xi$ values.
While the zero momentum quantities $\chi$ and $\chi_t$ show the expected
$z\approx 1$, the value for the energy seems to be even smaller. As $E$
can be written with a lattice derivative and couples to all nonzero
momenta, this behavior cannot be excluded for an integrated autocorrelation
time even if the slowest modes evolve with $z\approx 1$.
The analogous situation for free fields is discussed in \cite{UWOR}.
The topological charge was found to have very short
autocorrelations, which coincides with observations in \cite{HM2,IM}.

With the results on
exponents $z$ in \cite{HM1}, it is clear that for asymptotically
large correlation lengths multigrid becomes superior to OR. Based on
CPU times kindly communicated by the authors and on our own, we estimate the
crossover in efficiency to occur at $\xi \approx $ 20--30. In \cite{HM2}
a version of hybrid OR was tested which makes microcanonical moves in
the same embedded $U(1)$ spin models as are used for multigrid.
The crossover occurs at similar $\xi$, and exponents $z$ close to unity
are reached. It is interesting that the restriction of microcanonical
moves to a small submanifold of spin space seems to be of no disadvantage.
This opens the possibility of very simple exactly microcanonical
updates in complex field spaces like $SU(3)$.

Let us finally mention that the data shown here represent a total of
166 hours on one XMP processor, of which 2/3 went into the largest lattice.

\section{Results for the $CP^3$ model}

Apart from \cite{HM1,HM2} a number of other studies of
the $CP^3$ model have been attempted recently, among them \cite{JW,IM}.
They both used cluster algorithms that turned out not to be really efficient
enough to obtain high precision data at large correlation length.
The reason for the choice of $CP^3$ was in most cases
the expectation \cite{Lu}  that pathologies plague lattice definitions of
the topological charge for smaller $n$, at least when the standard action
and lattice are used.

Results from our runs on $L^2$ lattices are now summarized in the table:

\vspace{0.5cm}
\begin{center}
\begin{tabular}{|c|r|*{4}{r@{.}l|}}
\hline
$\beta$ &
\multicolumn{1}{c|}{$L$} &
\multicolumn{2}{c|}{$\chi$} &
\multicolumn{2}{c|}{$\chi_t \times 10^3$} &
\multicolumn{2}{c|}{$\xi_1$} &
\multicolumn{2}{c|}{$\xi_2$} \\ \hline
2.3 &   20 & 11&09(3) & 11&11(4) & 2&579(5) & 2&613(13) \\
2.5 &   32 & 26&17(5) &  5&20(2) & 4&442(7) & 4&488(16) \\
2.7 &   64 & 79&4(2)  &  1&731(8)& 8&800(21)& 8&845(46) \\
2.8 &   96 &145&9(5)  &  0&893(5)&12&640(36)&12&784(84) \\
2.9 &  128 &274&7(1.5)&  0&435(4)&18&49(9)  &18&79(22)  \\
3.1 &  256 &942&6(7.5)   &  0&103(1)&38&37(25) &39&32(59)  \\ \hline
\end{tabular}
\end{center}
\vspace{0.5cm}

All $1 \sigma$ errors quoted have been determined by a
 jackknife binning procedure
with various bin lengths and typically a few hundred
effectively independent bins. This has been compared with the
directly summed autocorrelation functions for $E, \chi, \chi_t$.
The two columns for the correlation length $\xi_{1,2}$ require some discussion.
We stored all estimates of the two point function at time separations
$t=0, \Delta, 2 \Delta, \ldots$ with $\Delta = L/32$ or $\Delta=1$ for
$L=20$. From two successive such values,
time dependent effective masses were determined.
The masses whose time separations selfconsistently embrace $j \xi$
are quoted as $\xi_j$. The table shows that $\xi_1$, for our statistical
accuracy, is systematically shorter than $\xi_2$.
This is the expected signature of higher mass states in the channel
probed by (\ref{C=}).
Inspection of $\xi_3$ suggests that, with the present level
of errors, $\xi_2$ is an acceptable estimate for the inverse massgap.
The ratio $\xi_3 / \xi_2$  is consistent with unity at the $ 1 \sigma $
level, although the mean exceeds unity for all but the last line in the table.
We face here the recurring dilemma of having to compromise
between systematic and
statistical errors when estimating energies.
As a consequence, it seems hard to exclude that under a worst case
scenario, with two nearby levels in the probed channel, we still underestimate
systematic errors. Comparing with cluster algorithms at this point,
we see that OR (and multigrid) is inferior not only in speed, but also the
lack of improved estimators is a severe disadvantage that is costly to overcome
with just statistics.
Finite size effects, on the other hand,
are expected to be irrelevant with $mL\approx 8$,
and this has been tested to some degree in \cite{HM1}.
Note that with the above procedure of extracting masses, and
for the chosen $\beta$ values,
all scales $(L, \xi, \Delta)$ of the various lattices are roughly in the
same proportion to one another. Therefore, neglecting scaling violations
due to the finite lattice spacing, the ratio $\xi_2/\xi_1$ should be
universal to a good approximation. Our data are consistent with this.
If we {\it assume} it, on the other hand,
we can determine the ratio to be about
1.01 from the smaller lattices, and apply this correction
to $\xi_1$ on the largest two lattices. This gives a smaller error
than taking $\xi_2$, and we feel
that the neglect of scaling violations for this 1\% correction is no problem.
We use these mass estimates in the following analysis.

\begin{figure}
\vspace{11cm}
\caption{Ratio of the mass to the two loop approximation of the
perturbative lattice scale}
\end{figure}
In \cite{HM1} asymptotic scaling was found to be violated in a way similar to
or even worse than for the  $O(n)$ models \cite{UWO348}.
We confirm this and plot in Fig.2 the ratio of the mass and the perturbative
 lattice scale in the two loop approximation
\be \label{LL}
\Lambda_L^{(2)} = (\pi \beta)^{1/2} \re^{-\pi\beta}.
\ee
Asymptotic scaling requires it to tend to a constant as
$\beta^{-1} \rightarrow 0$ with a speed linear in $\beta^{-1}$.
Clearly, no sign of this is visible in the (small) range over which we are
able to vary $\beta^{-1}$. We also consider a modified bare coupling as
discussed in \cite{UWO348}. From the perturbative expansion of the energy
\be
E= \langle \mbox{tr}[P(x) P(x+\muh)] \rangle = 1-\frac{3}{4}\beta^{-1}
-\frac{3}{16}\beta^{-2} +\mbox{O}(\beta^{-3})
\ee
we find that
\be
\beta_E=\frac{3}{4(1-E)}
\ee
is a correctly normalized alternative inverse bare (= short distance) coupling.
Now $\Lambda_L^{(2)}$ can be reexpanded in $\beta_E^{-1}$, and the series
is truncated to two loops in this parameter, whose values are taken from the
measured energies. This leads to the other data points in Fig.2. They
clearly  vary less steeply,
just like in the $O(n)$ case, but the behavior is not monotonic,
         and no really convincing extrapolation to $\beta_E^{-1}=0$ is feasible
either.
The difference between the two sets of data is best taken as evidence
of how far away we are from asymptotic behavior in the bare coupling.
In \cite{IM} a different lattice regularization of the $CP^3$ model has been
employed, and asymptotic scaling seems to be violated less drastically.
It would be very interesting to develop OR or multigrid for this action
and investigate universality for this case.

\begin{figure}
\vspace{11cm}
\caption{Scaling behavior of the topological susceptibility in
physical units}
\end{figure}
The situation with regard to scaling
is rather different for the dimensionless product of
physical quantities $\chi_t \xi^2$. In Fig.3 it is plotted against $1/\xi^2$,
the squared lattice spacing in physical units. We see that data from the
four largest lattices fit with a straight line and extrapolate to 0.156(2).
The intrinsic length scale for the scaling violations seems to be about
three lattice spacings, and for such correlation lengths their expansion
in powers of the lattice spacing ceases to be useful. The breakaway is
rather abrupt. While for $\xi=8.8$ the leading term alone works very well,
each of the smaller lattices needs a new power of $\xi^{-2}$ to be
included to get an acceptable fit. We wish to remark that an asymptotic
expansion in powers $\xi^{-2}$ (neglecting the variation of logarithms)
is suggested by renormalized perturbation theory \cite{Sym} and by
an effective action picture \cite{Pol}. From the point of view of statistical
mechanics, an arbitrary power of $\xi^{-1}$ would seem natural.
We found that the four leftmost points in Fig.3 can indeed be extrapolated
with anything between $\xi^{-1}$ to $\xi^{-4}$. As our data get very close
to the continuum in Fig.3 (compare this with Fig.2!), the extrapolated values
are not too different. If one wants to include these uncertainties in a
systematic error one could still quote $\chi_t/m^2 = 0.16(1)$ in the
continuum.

In \cite{HM2} $\chi_t \xi^2$ is found to fall again at $\beta=3.1,3.3$
after rising and agreeing with our data for smaller $\beta$ values. Since we
do not have data for $\beta=3.3$, the only real discrepancy
in the measured data hinges upon
a smaller value beyond errors for $\xi$ at $\beta=3.1$ in \cite{HM2}.
As mentioned in \cite{HM1} the authors extract correlation lengths by
fitting over a range of separations starting at $\xi$. We feel that there
may also be a problem with higher states and propose a simple analysis
with effective masses. In any case this discrepancy should be resolved
in the future, if necessary, with new long runs.

Let us finally compare with results from the large $n$ expansion \cite{Lu}.
To leading order one expects
\be
\chi_t \xi^2 = \frac{3}{4 \pi n} = 0.05968\ldots \mbox{ for } n=4,
\ee
from which we differ by a factor of about 2.6.
First $1/n$ corrections have been computed in \cite{ln1}.
The authors find
\be
\chi_t \tilde{\xi}^2 = \frac{1}{2 \pi n} \left(1-\frac{0.38}{n}\right)
= 0.03979 (1-0.095) \quad \mbox{ for }n=4,
\ee
where $\tilde{\xi}$ is a second moment definition
of the correlation length. The authors argue that, although
$\tilde{\xi}^2=2 \xi^2/3$ holds at $n=\infty$, the two lengths differ
only by a few \% for small $n$ with the connection being nonanalytic
in $1/n$. This leads to an even larger discrepancy between our result
and large $n$ despite an already small leading correction for $n=4$.
In summary, no quantitative agreement with the $1/n$ expansion at $n=4$
is found.

\noindent
{\large\bf Acknowledgements} \newline
The author would like to thank M. Campostrini, M. Hasenbusch and S. Meyer for
helpful correspondence.

\end{document}